\documentclass[twocolumn,showpacs,amsmath,amssymb]{revtex4}
\usepackage{graphicx}
\usepackage{dcolumn}
\usepackage{bm}
\begin{document}
\preprint{APS/*!*!*}
\title{Effective mass enhancement in two-dimensional electron systems:
the role of interaction and disorder effects}

\author{R. Asgari and B. Davoudi}
\affiliation{NEST-INFM and Classe di Scienze, Scuola Normale Superiore, 
I-56126 Pisa, Italy\\
and\\
Institute for Studies in Theoretical Physics and Mathematics, Tehran
19395-5531, Iran\\}
\author{B. Tanatar}
\affiliation{Department of Physics, Bilkent University, 06800 Bilkent, 
Ankara, Turkey}
\date{\today}
\begin{abstract}
Recent experiments on two-dimensional (2D) electron systems have found 
a sharp increase in the effective mass of electrons with decreasing 
electron density. In an effort to understand this behavior we employ
the many-body theory to calculate the quasiparticle effective mass
in 2D electron systems. Because the low density regime is explored in
the experiments we use the $GW\Gamma$ approximation where the vertex 
correction $\Gamma$ describes the correlation effects to calculate the 
self-energy from which the effective mass is obtained. 
We find that the quasiparticle effective mass shows a sharp
increase with decreasing electron density. Disorder effects
due to charged impurity scattering plays a crucial role in
density dependence of effective mass.
\end{abstract}

\maketitle

There has been a large amount of experimental and theoretical activity 
in recent years to understand the ground state properties of 
homogeneous two-dimensional (2D) electron systems. Advances in 
fabrication techniques have made it possible to probe various 
quantities of interest in high quality and very low density samples. 
Most notably, the observation of a metal-insulator transition\cite{1} 
in these systems provides a major motivation to study the various
physical properties.
In recent experiments the spin susceptibility, Lande $g$-factor, and 
effective mass are measured for 2D electron systems made of Si-MOSFETS 
and GaAs quantum-well structures.\cite{2,3,4,5,6,7} In particular, 
Shashkin {\it et al}.\cite{3,4} reported a sharp increase of effective 
mass near the critical density at which the system starts to show 
deviations from the metallic behavior. On the other hand, Pudalov {\it
et al}.\cite{2} have found only moderate enhancement of the spin
susceptibility and effective mass in their samples.

There has been a number of calculations of the 
quasiparticle properties including the effective mass of 2D electron 
gas employing a variety of approximations.\cite{8}
More recent theoretical calculations of the effective mass of 2D
electrons concentrated on the density, spin polarization and 
temperature dependence.\cite{9,10,11,12,13} 

In view of the different experimental results and their controversial
interpretation we have addressed in this work the density dependence 
of the effective mass in an interacting electron system  at $T=0$
in the presence of charged impurities.
We employ the $GW\Gamma$ approximation\cite{14} which includes the 
vertex corrections in an approximate way to calculate the quasiparticle
effective mass. The local-field factor describing the correlation
effects which enters the vertex function
$\Gamma$ is obtained within the memory function formalism and the
self-consistent field method. We had previously shown\cite{15} that 
such an approach correctly describes the anomalous behavior of the
thermodynamic compressibility of 2D electron systems as reported
experimentally.\cite{16} In the present work, we extend our earlier
considerations to calculate the effective mass. We find that the 
quasiparticle effective mass is greatly enhanced at low density in the 
same region when the compressibility diverges. 

In the following we first outline the theoretical framework
with which we calculate the quasiparticle effective mass of 2D
electron system. We next present our results within various 
levels of approximations. We discuss the results of our calculations
in view of other theoretical approaches and experimental findings.
We conclude with a brief summary.

We consider a 2D electron system interacting via the long range Coulomb
interaction $V_q=2\pi e^2/(\epsilon_0 q)$ where $\epsilon_0$ is the 
background dielectric constant. The system is characterized by the 
dimensionless interaction strength $r_s=1/(\pi n {a_B^{\ast}}^2)^{1/2}$,
where $n$ is the 2D electron density and $a_B^{\ast}=
\hbar^2\epsilon_0/(m e^2)$ is the effective Bohr radius defined 
in terms of the band mass $m$ of electrons in the semiconductor 
structure. 

We use the theoretical framework developed by Thakur {\it et
al}.\cite{17} employing the memory-function formalism and the
self-consistent field method to calculate the density-density response
function of a disordered electron system.
The effect of disorder is to dampen the charge-density fluctuations 
and modify the response function. In a number-conserving approximation 
the density-density response function for noninteracting electrons 
is given by
\begin{equation}
\chi_0(q,\omega;\gamma)={\chi_0(q,\omega+i\gamma)\over
1-{i\gamma\over\omega+i\gamma}\left[1-{\chi_0(q,\omega+i\gamma)\over
\chi_0(q)}\right]}
\end{equation}
where $\gamma$ is the scattering rate. The correlation effects are
described by the generalized random-phase approximation (RPA) for the
interacting system density-density correlation function
\begin{equation}
\chi(q,\omega;\gamma)={\chi_0(q,\omega;\gamma)\over
1-V_q[1-G(q)]\chi_0(q,\omega;\gamma)}
\end{equation}
in which the static local-field factor $G(q)$ embodies the correlation
effects. In this work we use the self-consistent field method of Singwi
{\it et al}.\cite{18} to calculate $G(q)$. As a simplified model
we also consider the Hubbard local-field factor given as
$G(q)=q/2\sqrt{q^2+k_F^2}$.

Within the memory function formalism the scattering rate is expressed 
in terms of the screened disorder potential and the relaxation 
function as\cite{17,19}
\begin{equation}
i\gamma=-{n_i\over 2m n}\sum_q q^2 {\langle |U_{\rm
imp}(q)|^2\rangle\over \varepsilon^2(q)}
{\phi_0(q,i\gamma)\over 1+i\gamma{\phi_0(q,i\gamma)\over
\chi_0(q)}}
\end{equation}
where $U_{\rm imp}(q)=V_q e^{-qd}$ is the impurity potential for 
charged impurities located at a distance $d$ away from the 2D electron 
layer, and $n_i$ is the impurity concentration.
The relaxation function is given by\cite{17,19}
$\phi_0(q,i\gamma)=[\chi(q,i\gamma)-\chi_0(q)]/(i\gamma)$ and 
$\varepsilon(q)=1-V_q[1-G(q)]\chi_0(q)$ is the static screening
function. Because the scattering rate $\gamma$ depends on the screening
function $\varepsilon(q)$ which itself is determined by the disorder
included response function the above set of equations are solved
self-consistently.

The quasiparticle properties of the 2D electron system are obtained 
from the self-energy function\cite{14} $\Sigma(k,\omega)$ which we
calculate at zero temperature. Since we are
interested in exploring the interaction effects we include the vertex
corrections to the self-energy and employ the $GW\Gamma$
approximation.\cite{14}
The self-energy in the $GW\Gamma$ approximation is 
written as a sum of two terms
$\Sigma(k,\omega)=\Sigma^{\rm line}(k,\omega)+
\Sigma^{\rm pole}(k,\omega)$
where
\begin{equation}
\Sigma^{\rm line}(k,\omega)=-\sum_q
V_q\,\int_{-\infty}^{\infty}{d\omega'\over
2\pi}\,{\Gamma(q,i\omega')\over\varepsilon(q,i\omega')}\,
{1\over\omega+i\omega'-\xi_{k+q}}
\end{equation}
and
\begin{equation}
\Sigma^{\rm pole}(k,\omega)=\sum_q
V_q[\theta(\omega-\xi_{k+q})-\theta(-\xi_{k+q})]\,
{\Gamma(q,\xi_{k+q}-\omega)\over
\varepsilon(q,\xi_{k+q}-\omega)}
\end{equation}
in which $\xi_k=k^2/2m-E_F$ is the single-particle energy measured 
relative to the Fermi energy. The vertex function in the
local-approximation is given as\cite{14}
\begin{equation}
\Gamma(q,\omega)={1\over 1+V_qG(q)\chi_0(q,\omega;\gamma)}
\end{equation} 
in terms of the local-field factor $G(q)$ describing correlation
effects beyond the RPA. 
The dielectric function appearing in Eqs.\,(4) and (5) is given by
$\varepsilon(q,\omega)=1-V_q \chi_o(q,\omega;\gamma)\Gamma(q,\omega)$. 
The above expressions for the self-energy reduce to the $GW$-RPA
results when we set $G(q)=0$. Furthermore, taking $\gamma=0$, we recover
the results for a clean system.

We use the on-shell approximation to the self-energy in the single
particle spectrum $E_k=\xi_k+\Sigma(k,\xi_k)$ to obtain the effective 
mass perturbatively\cite{8,11,14}
\begin{equation}
{m^{\ast}\over m}=\left[1+{\partial\Sigma\over\partial\omega}
+\frac{m}{k} {\partial\Sigma\over\partial k}\right]^{-1}\, ,
\end{equation}
where the frequency and momentum derivatives of 
$\Sigma(k,\omega)$ are evaluated at the Fermi surface. Such an approach
is argued to be more appropriate over solving the full Dyson's equation
$E_k=\xi_k+\Sigma(k,E_k)$, since the self-energy is calculated
using the noninteracting Green function. The resulting scheme 
incorporates the higher order diagram contributions 
better.\cite{8,11,14}

In the numerical calculations we specialize to GaAs systems for which
some measurements of the effective mass are undertaken.\cite{6} Since 
the dominant scattering mechanism is known to be that due to the 
charged impurities we take $d=250$\,\AA\ for the setback distance in 
$U_{\rm imp}(q)$, and consider $n_i$ to be of the order of $\sim
10^{10}$\,cm$^{-2}$. We first solve the self-consistent equations for 
the scattering rate $\gamma$ and the local-field factor $G(q)$ as a 
function of $r_s$ and impurity density $n_i$. These quantities 
determine the dynamic screening function $\varepsilon(q,\omega)$ and 
the vertex function $\Gamma(q,\omega)$. Afterward the self-energy and 
its derivatives are evaluated to find the effective mass $m^{\ast}/m$.

\begin{figure}
\includegraphics[scale=0.5]{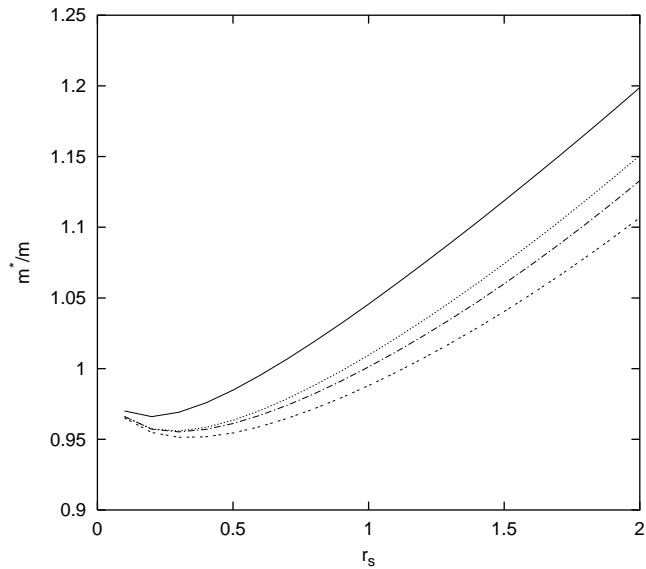}
\caption{The quasiparticle effective mass as a function of $r_s$
in the range $0<r_s<2$. The solid and dot-dashed lines
indicate the $GW$-RPA and $GW\Gamma$ approximations, respectively.
$GW\Gamma$ approximation which uses the Hubbard local-field factor is
indicated by the dotted line. $GW\Gamma$ approximation including the
charged impurity scattering (with impurity concentration $n_i=0.5\times
10^{10}$\,cm$^{-2}$) is shown by the dashed line.}
\end{figure}

In Fig.\,1 we show the effective mass $m^{\ast}/m$ as a function of 
$r_s$ calculated in various theoretical approaches. 
We observe that even at these
relatively higher densities there are notable differences between
the $GW$ and $GW\Gamma$ approximations indicating the importance of
vertex corrections. It appears that the correlation effects suppress
the effective mass renormalization. The disorder effects due to 
charged impurity scattering tend to increase the effective mass with 
respect to the clean system. There is still a significant reduction in 
$m^{\ast}/m$, however, compared to the $GW$-RPA result.

\begin{figure}
\includegraphics[scale=0.5]{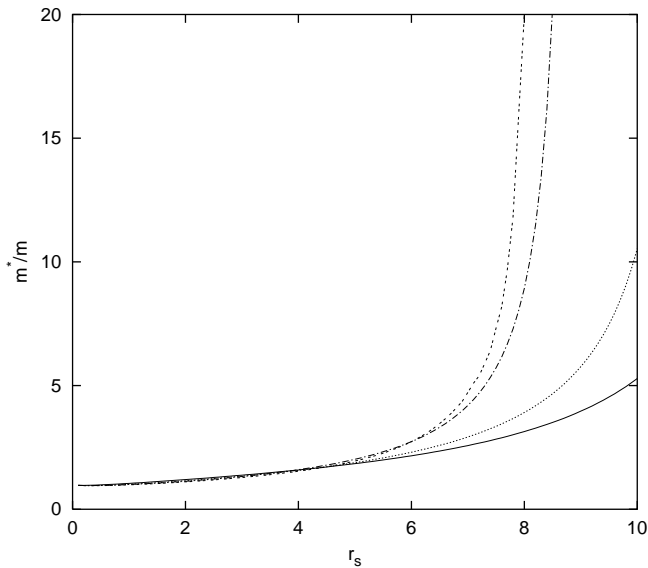}
\caption{The quasiparticle effective mass as a function of $r_s$
in the range $0<r_s<10$. 
The solid and dot-dashed lines
indicate the $GW$-RPA and $GW\Gamma$ approximations, respectively.
$GW\Gamma$ approximation which uses the Hubbard local-field factor is
indicated by the dotted line. $GW\Gamma$ approximation including the
charged impurity scattering (with impurity concentration $n_i=0.5\times
10^{10}$\,cm$^{-2}$) is shown by the dashed line.}
\end{figure}

We next display the quasiparticle effective mass at much lower
densities. Strictly speaking, $GW$ approximation is
valid only in the high density limit, inclusion of the vertex
corrections in $GW\Gamma$ approximation improves the regime of validity.
In any case, we wish to explore the effective mass trends in the low 
density, strongly interacting region.
Figure 2 shows $m^{\ast}/m$ calculated 
in various theoretical approaches. $GW$-RPA yields a modest
enhancement for the whole density range. The suppression found in 
Fig.\,1 for the $GW\Gamma$ approximation reverses its behavior around
$r_s\approx 4$ and shows an enhancement relative to the $GW$-RPA
results at lower densities. Qualitatively similar behavior is
obtained when we use the simple Hubbard local-field factor
within the $GW\Gamma$ approximation. 
A notable feature of the $GW\Gamma$ approximation results is that
effective mass exhibits a sharp increase around $r_s\sim 8$. 
That the strong interaction effects would lead to a large enhancement
in $m^{\ast}/m$ is also evident when the Hubbard local-field factor
is used within the $GW\Gamma$ approximation. Finally, when charged
impurity scattering effects are included in the calculation
we find that a similar sharp increase in $m^{\ast}/m$ occurs at a
smaller $r_s$ value. We have also calculated the effect of impurity
scattering for different parameter values of $d$ and $n_i$ and found
qualitatively similar results.

Although the results of $GW$ and $GW\Gamma$ approximations at large
$r_s$ are should be taken in with caution, the low density trends of
$m^{\ast}/m$ should be indicative.
In this perspective our calculations indicate that the effective mass
enhancement in 2D electron systems can be accommodated within the Fermi
liquid theory when the vertex corrections describing the strong 
correlation effects are taken into account. 
In particular, a sharp increase in $m^{\ast}/m$ as shown in Fig.\,2
is quite suggestive in view of the recent experimental 
findings.\cite{3,4} Effective mass enhancement is also observed in 2D
neutral Fermi systems\cite{20} and represented by a $GW$-type
calculation.\cite{21} 
Our calculations also show that the $r_s$ value at which $m^{\ast}/m$ 
exhibits a sharp increase can be controlled by disorder effects. 
In a self-consistent scheme where remote charged impurities are taken 
into account we find that the large enhancement in $m^{\ast}/m$ occurs 
at a higher density compared to the strongly interacting clean system.

Recent theoretical approaches\cite{9,10} have modeled the low density 
electron liquid as close to the Wigner crystallization to obtain a 
strong increase in the effective mass.
On the other hand, Morawetz\cite{12} found a divergent behavior in
$m^{\ast}/m$ at the metal-insulator transition by considering the
scattering from heavy impurity ions, and Galitski and Khodel\cite{13} 
attribute divergence of the effective mass to the density wave 
instability.
Our theoretical scheme considers
the metallic regime, therefore we cannot distinguish the nature of
the possible new state beyond the critical $r_s$ value.
However, our earlier calculations\cite{15} of the anomalous behavior
of the compressibility at around the same range of $r_s$ values points
to a possible metal-insulator transition approaching from the metallic 
side.
A related quantity of interest would be the spin susceptibility or the
$g$ factor for which experimental results are available.
These quantities require the calculation of density-density response
function and the local-field factor as functions of the spin 
polarization which were not undertaken in this work. 

Our calculations were performed at $T=0$ and for zero thickness 2D
electron layers. It would be interesting to
extend our work to finite temperatures and to finite width quantum
wells to make better contact with experiments. 
As the Coulomb interaction effects will be less strong in quantum 
wells, it is expected that the enhancement of 
$m^{\ast}/m$ will be less marked. 

In summary, within a many-body approach which takes the
electron-electron and electron-impurity interaction effects into account
we have calculated the effective mass of a 2D electron system at zero
temperature.
We have found within the commonly used on-shell approximation that 
$m^{\ast}/m$ is highly enhanced at larger values of 
the density parameter $r_s$ as a result mainly of the correlation 
effects. The interplay between the correlation 
effects and the impurity scattering influences qualitative changes 
in this behavior. Our comparative results should provide insight into
the workings of many-body methods in the strong interaction regime.\\

This work was partially supported by MIUR under the PRIN2001 Initiative.
B.\,T. acknowledges the support by TUBITAK, NATO-SfP, MSB-KOBRA001, and 
TUBA. 


\end{document}